\documentclass[%
 aps,
 sd,%
 amsmath,amssymb,reprint,
jmp]{revtex4-1}%
\usepackage[pdftex]{graphicx}
\usepackage{dcolumn}
\usepackage{bm}
\usepackage{color,soul}
\usepackage{float}

\newcommand{\vect}[1]{\bm{#1}}

\begin{document}

\title{Transient dynamics of nonlinear magneto-optical rotation}
\author{R. S. Grewal}
\affiliation{Institute of Physics, Jagiellonian University, \L ojasiewicza 11, 30-348 Krak\'ow, Poland}
\author{A. Rybak}
\affiliation{ABB Corporate Research Center, Starowi\'slna 13A, 31-058, Krak\'ow, Poland}
\author{M. Florkowski}
\affiliation{ABB Corporate Research Center, Starowi\'slna 13A, 31-058, Krak\'ow, Poland}
\author{S. Pustelny}
\email{pustelny@uj.edu.pl}
\affiliation{Institute of Physics, Jagiellonian University, \L ojasiewicza 11, 30-348 Krak\'ow, Poland}

\begin{abstract}

We analyze nonlinear magneto-optical rotation (NMOR) in rubidium vapor subjected to continuously-scanned magnetic field. By varying magnetic-field sweep rate, a transition from traditionally-observed dispersive-like NMOR signals (low sweep rate) to oscillating signals (higher sweep rates) is demonstrated. The transient oscillatory behavior is studied versus light and magnetic-field parameters, revealing a strong dependence of the signals on magnetic-sweep rate and light intensity. The experimental results are supported with density-matrix calculations, which enable quantitative analysis of the effect. Fitting of the signals simulated versus different parameters with a theoretically-motivated curve reveals presence of oscillatory and static components in the signals. The components depend differently on the system parameters, which suggests their distinct nature. The investigations provide insight into dynamics of ground-state coherence generation and enable application of NMOR in detection of transient spin couplings.
  
\end{abstract}

\maketitle
\section{Introduction}

Light, coherently coupling energy states of a microscopic system (e.g., atom, molecule, ion, etc.), may induce quantum coherence between the states. In a specific example of interaction of atoms with not-too-weak linearly ($\sigma$) polarized light, in which at least one of the coupled levels has a rich magnetic structure (numerous magnetic sublevels), Zeeman coherences between the sublevels are generated. This may result in  pumping of the atoms into a coupled state (dark state), where they do not absorb resonant light \citep{Auzinsh2009Optically}. More generally, the coherence may give rise to various nonlinear optical phenomena, many of which currently are objects of intense studies \citep{Li95,ChenEIT,Valente,Pack,NikolicEIT,RenzoniCPT,SihongCPT,Margalit1,Margalit,Momeen,Pustelny2011Tailoring}.  

The evolution of atomic polarization/quantum coherences has been studied in both  transient and steady states in various light-induced coherent phenomena. In the context of transient behavior, such effects as electromagnetically induced transparency (EIT) and electromagnetically induced absorption (EIA) \citep{Li95,ChenEIT,Valente,Pack,NikolicEIT}, coherent population trapping (CPT) \citep{RenzoniCPT,SihongCPT,Margalit1,Margalit}, and nonlinear magneto-optical rotation (NMOR) \citep{Momeen,Pustelny2011Tailoring} were studied. In most of these studies, the transient response of the atomic medium were examined either by sudden change of light frequency or magnetic-field strength \citep{Valente,SihongCPT}. For example, Valente \textit{et al.} studied the temporal evolutions of EIT and EIA resonances in an Rb vapor as the magnetic field was suddenly turned on or off \citep{Valente}. The authors observed that when the field was suddenly switched on, the transient oscillations of optical signals are observed in both EIT and EIA resonances. At low light intensity, the dynamics of the oscillations in EIT and EIA was similar, i.e., same decay rates of the oscillations were observed. At high light intensity, the EIA oscillations decayed faster as EIT ones. This originated from different dynamics of coherence generation and relaxation in both cases. Park \textit{et al.} investigated the transient oscillations of CPT signals when the laser frequency was changed by continuous sweeping (triangular waveform) or sudden detuning of light \citep{Park}. When light was abruptly detuned from the optical resonance the oscillations were observed. It was shown that the time period of the oscillations was inversely proportional to the  detuning. Similar  results were obtained in EIT spectra of laser-cooled ${}^{87}{\rm{Rb}}$ atoms produced by a magneto-optical trap \citep{Chen}.  

The transient dynamics of optical signals can also be used practically. 
For example, atomic clock based on transient oscillations of CPT signal were demonstrated \citep{Guo,Wang,Dawei}.
 Recently, the time domain analysis of the coherent transient oscillations has been used for magnetometry \citep{Lenci1,Behbood,Lenci,Breschi2014,Quan2017}. In these studies, first the atomic polarization was generated in a medium through optical pumping. Next, an auxiliary magnetic field was switched off and the precession of atomic alignment due to the field was measured as a function of time. Finally, the Larmor frequency corresponding to unknown magnetic field was directly extracted by numerically fitting of measured transmission signal and hence the field strength was determined \citep{Lenci1,Lenci,Quan2017}.

The transient response of polarization rotation due to continuous change of the magnetic field in a Rb-vapor cell was examined by Momeen \textit{et al.} \citep{Momeen}. These studies were performed at high laser power, so that the excited-state population of the atomic transitions was saturated. The authors experimentally demonstrated the oscillatory bahavior of NMOR signal. They showed that the dynamics of the signal depends on the hyperfine level light is tuned to. Particularly, at low magnetic-field sweep rate, the linewidth of a resonance observed with light tuned to a lower ground-state hyperfine level was narrower than that observed for upper ground-state hyperfine level tuning. Dimitrijevi\'c \textit{et al.} \citep{Dimitrijevic} theoretically analyzed the transient behavior of NMOR of light pulse, propagating through a cold atomic gas. It was shown that during rising pulse intensity a traditional, a dispersively-shaped NMOR signal was observed and during its falling oscillatory behavior of rotation angle was observed. 

In the present work, a detailed study on the transient response of NMOR signals is performed. We analyze the transient behavior dependence on the different parameters, including light intensity and magnetic-field amplitude and modulation frequency (sweep rate). First, the magnetic-field is swept at low sweep rate and a situation similar to that encounter in traditional NMOR is reproduced. Later, the faster sweep rates are investigated. In the case, the oscillations of polarization rotation are detected while scanning magnetic field across zero. We demonstrate that it is the sweep rate not amplitude or frequency of the modulation that determines the system dynamics. We also investigate a role of pumping and relaxation rate on system’s transient dynamics via measurements of NMOR signals at different light intensities. We show that increasing of light intensity leads to faster decay of the oscillations, and at higher intensities, all the oscillations are completely deteriorated. We analyze the signals quantitatively by fitting them with theoretically motivated curve. The analysis shows that the amplitude of the oscillations decreases as the sweep rate is increased but their decay rate remains same. All these results are confirmed with density-matrix calculations. 

\section{Experimental setup}
The schematic diagram of experimental setup is shown in Fig.~\ref{setup}(a). A paraffin-coated buffer-gas-free vapor cell, containing isotopically enriched sample of $^{87}{\rm{Rb}}$, is placed at the center of four-layer magnetic shield (three mu-metal layers; one innermost ferrite layer). The shield attenuates external magnetic field by at least a factor of ${10}^6$ \citep{TwinLeafShield}. The cell is operated at room temperature, corresponding vapour density of about ${10}^9$~\rm{atoms/cm$^3$. Magnetic-field coils [not shown in Fig.~\ref{setup}(a)] are installed inside the shield to further compensate the residual magnetic fields 
 and to generate magnetic field along the light propagation direction. A computer-controlled multi-channel current source is used to drive current through the coils. The longitudinal magnetic field is generate and scanned with a function generator. The field is scanned at different sweep rates by varying scanning frequency (between 10~mHz and 20~Hz) and magnetic-field amplitude (between 0.14~mG to 1~mG). 
 
\onecolumngrid

 \begin{figure}[H]
\centering  
\includegraphics[height=7.0cm,width=15.8cm]{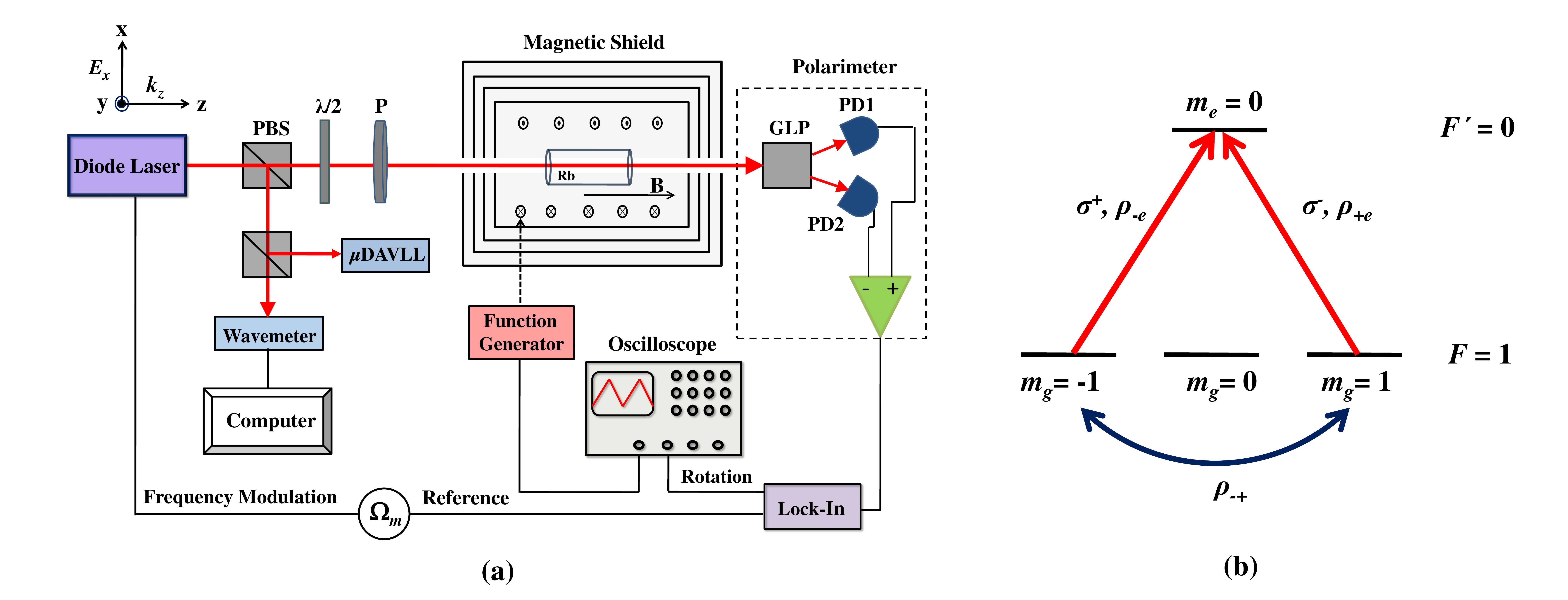}
\caption{(a) Schematic diagram of  experimental setup used to measure NMOR signals. PBS is the  polarizing beam splitter; $\lambda /2$ is the half-wave plate; P is the polarizer; GLP is the Glan-laser crystal polarizer; PD1 and PD2 are the photo-detectors; $\mu$DAVLL is the dichroic atomic vapor laser lock system, exploiting a micro-size rubidium vapor cell. (b) Two-level atomic system $F = 1 \rightarrow F' = 0$ interacts with left ($\sigma^-$) and right ($\sigma^+$) circular polarized components of the linear polarized light in  the $\Lambda$-configuration. Optical coherences between ground-state and excited-state Zeeman sublevels are given by $\rho_{-e}$ and $\rho_{+e}$, and $\rho_{-+}$ represents the ground-state Zeeman coherence.}
\label{setup}
\end{figure}
\twocolumngrid

A diode laser (Toptica DL Pro), emitting light of a wavelength of 795 nm and spectral width of less than 1 MHz, is used. The light propagtes along \textit{z}-axis ($\vect{k}\|\vect{z}$) with  polarization vector along the \textit{x}-axis ($\vect{E}\|\vect{x}$). The laser frequency is modulated at  $\Omega_m/(2\pi)=80$~kHz with a modulation amplitude of a few hundred megahertz. Application of the frequency-modulated light is not motivated by the desire of detection of stronger magnetic field, such as in Ref.~\citep{Budker2002FNMOR}, but rather the willingness of application of phase-sensitive detection and increasing signal-to-noise ratio. The laser is tuned to the low-frequency wing of the Doppler broadened $F_{g}=2 \to F_{e}=1$ transition of $^{87}{\rm{Rb }}$ D$_{1}$ line, where maximum rotation signal is observed, and stabilized with a dichroic atomic vapor laser locked exploiting micro-cell filled with rubidium vapor and buffer gas \citep{Pustelny2016}. The frequency of the laser is monitored with the saturation absorption spectroscopy system and HighFinesse WS Ultimate wavemeter. 

Rotation of a polarization plane of linearly polarized light is detected using a balanced polarimeter, consisting of a Glan-laser crystal polarizer (GLP), which axis is oriented at $45^\circ$ with respect to a polarizer (P) placed before the vapor cell, and two photodiodes, measuring intensity of light directed into two channels of GLP. The photodiode difference signal is demodulated with a lockin amplifier, operating at the first harmonic of the light's modulation frequency. High modulation frequency enables application of a short time constant (300~$\mu$s), which ensures absence of signal distortion due to phase-sensitive detection. The locking output signal is observed with a digital scope. The measurements are performed at various light intensities, controlled using a half-wave plate ($\lambda/2$) situated before the input polarizer P, and sweep rates. 

\section{THEORETICAL MODELING\label{sec:Modeling}}

A theoretical analysis using a density-matrix approach is performed to understand a quantum evolution of the system. The time evolution of the density matrix $\rho$ is given by the Liouville equation 
\begin{equation}
\dot \rho  =  - \frac{i}{h}\left[ {H,\rho } \right] - \frac{1}{2}\left\{ {\Gamma ,\rho } \right\} + \Lambda ,
\label{liou}
\end{equation}
where $H$ is the total Hamiltonian incorporating such processes as optical pumping and interaction of atoms with external magnetic field. Spontaneous decay of the excited state and uniform relaxation of all atomic states is described by the relaxation operator $\Gamma$. Repopulation of the system due to reservoir is described with the operators $\Lambda$. The square and curly brackets represent the commutation and anticommuniation operations, respectively.

The total Hamiltonian $H$ can be written as the sum of an unperturbed Hamiltonian $H_{0}$ and the operators describing interaction of atoms with light $V_l$ and magnetic field $V_B$. Here, we consider the light-atom interaction in the dipole approximation $V_l$
\begin{equation}
	V_l=-\vect{d}\cdot\vect{E}=-\frac{d_xE_0}{2\sqrt{2}}\left(e^{i\omega t}+e^{-i\omega t}\right)\!\!,
	\label{eq:Vl}
\end{equation}
where $\vect{d}$ is the electric dipole momentum and $d_x=qx$ is the $x$-component of the operator, with $q$ being the elementary charge and $x$ being the $x$-position operator. $\vect{E}$ is the electric field of $x$-polarized light which is polarized along \textit{x} direction, $E_0$ is the amplitude of the electric field of light, and $\omega$ is the light frequency. The time-dependence of the optical frequency is eliminated from the description by application of rotating-wave approximation \citep{Sudyka2017RWA}. 

The magnetic-field interaction operator may be written as 
\begin{equation}
	V_B=-\vect{\mu}\cdot\vect{B}=-\mu_B g_FF_zB(t),
	\label{eq:VB}
\end{equation}
where $\vect{\mu}$ is the magnetic dipole moment, $\mu_B$ is the Bohr magneton, $g_F$ is the Land\'e factor, $F_z$ is the $z$-projection spin operator, and $B(t)$ is the time-varying magnetic-field amplitude. The time-dependent magnetic field,  oriented along the light propagation direction,  is modulated with triangular waveform
\begin{equation}
	\bm{B}(t)=\frac{2B_0}{\pi}\arcsin\left[\sin\left(\Omega_n t\right)\right]\vect{z},
\end{equation}
where $\Omega_n$ and $B_0$ are the magnetic-field modulation frequency and amplitude, respectively, and $\vect{z}$ is the unit vector. 

In our simulations, a two-level system, with $F=1$ in the ground state and $F' = 0$ in the excited state, is considered  [Fig.~\ref{setup}(b)]. In the scheme, the $x$-polarized light, consisting of $\sigma ^\pm $ components,  couples the ground-state Zeeman sublevels $m_g=\pm 1$ with the excited-state sublevel $m_e= 0$. This generates optical coherences between ground and excited states ($\rho_{-e}$ and $\rho_{+e}$) but also the Zeeman coherence $\rho_{-+}$ between the $m_F=1$ and $m_F=-1$ ground-state magnetic sublevels.

The polarization rotation is determined by the optical coherences (see, for example, Ref.~\citep{Pustelny2015Nonlinear} Supplemental Material)
\begin{equation}
	\varphi\propto \textrm{Re}\left(\sigma_{-e}-\sigma_{+e}\right),
\end{equation}
where $\sigma_{\pm e}$ are slowly evolving envelops of the optical coherences $\rho_{\pm e}$, $\rho_{\pm e}=\sigma_{\pm e}\exp(-i\omega t)$. As shown, for example, in Ref.~\citep{Pustelny2011Tailoring}, the optical coherences depend on the ground-state Zeeman coherences [curved arrow in Fig.~\ref{setup}(b)], so that the narrowest component (with respect to the magnetic field) of the observed NMOR signal is determined by the ground-state coherence
\begin{equation}
	\varphi\propto \textrm{Re}\left(\rho_{-+}\right).
\end{equation}

Substituting Eqs.~\eqref{eq:Vl}-\eqref{eq:VB} into Eq.~\eqref{liou}, one gets the formula for the time evolution of the Zeeman coherences in the system
\begin{equation}
{\dot \rho _{_{ -  + }}} =  - \gamma {\rho _{ -  + }} + i\frac{{{\Omega _R}}}{{2\sqrt 6 }}({\rho _{ - e}} + {\rho _{e + }}) 
+ i\frac{{4{\mu _B}{g_F}{B_0}}}{\pi }{\kern 1pt} {\rm{arcsin}}[{\rm{sin}}({\Omega _n}t)]\,{\rho _{ -  + }}
\label{eq:ZeemanCoherence}
\end{equation}
where ${\Omega _R} = \frac{{{d_x}{E_0}}}{\hbar }$ is the Rabi frequency of  light and $\gamma$ is the ground-state relaxation rate.

Numerical calculations of Eq.~\eqref{eq:ZeemanCoherence} enable simulations of dynamic of NMOR signal for modulated magnetic field. Below, the results of these simulations are compared with the experimental data.

\section{Results and discussion\label{sec:Results}}

Figure~\ref{diffSweepRate} shows NMOR signals measured as a function of time at different magnetic-field sweep rates. For low sweep rates ($\lesssim 10$~$\mu$G/s), the signal is similar to a traditionally recorded NMOR signal [Fig.~\ref{diffSweepRate}(a)]. However, the more thorough analysis reveals a small asymmetry between two sides of the signal (one arm of the resonance is larger than the other). This is not observed in conventional NMOR signals and indicates in-complete equilibration of the system in Fig. 2(a). It was verified with independent measurements that the asymmetry disappears for even lower sweep rates (not shown).
\onecolumngrid
 
\begin{figure}[H]
\centering 
	\includegraphics[width=15.6cm]{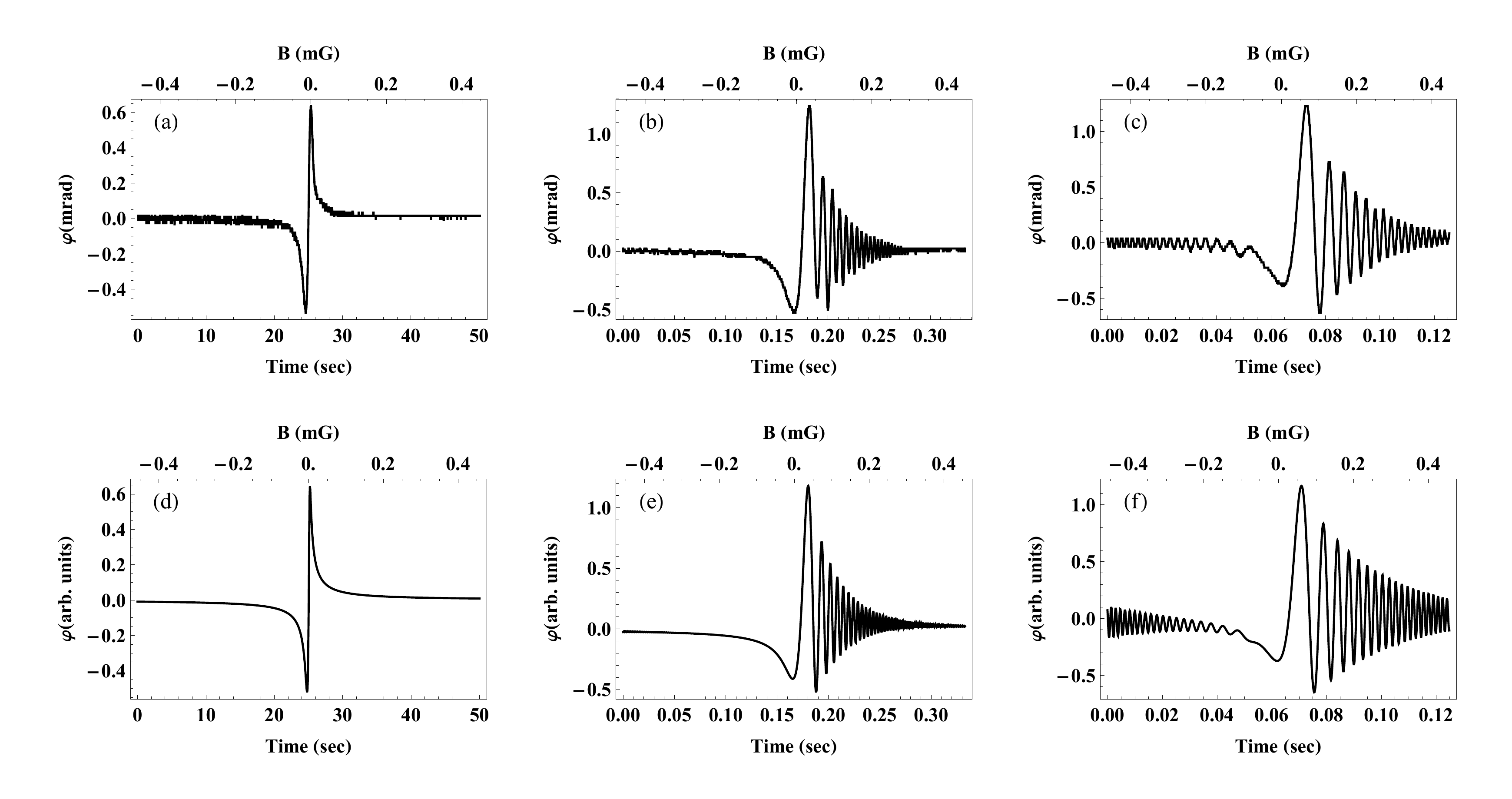}
	\caption{Experimentally measured (top row) and numerically simulated (lower row) NMOR signals as a function of time and magnetic field at different magnetic field sweep rates. For small sweep rates, $4.5$~$\mu$G/s in (a) and (d), the signals are quite similar to a conventional NMOR signal (with a small asymmetry between two wings). For larger rates, $700$~$\mu$G/s in (b) and (e),  $1800$~$\mu$G/s in (c) and (f), the oscillations of polarization rotation are observed while crossing $B=0$. The period of oscillations varies with time due to the change of magnetic field. The laser beam intensity is set to 2 $\rm{\mu W/mm^2}$ at the cell. The parameters used in simulations are, ${\Omega _R}/2\pi = 10\,{\rm{kHz}}$, $\gamma/2\pi  = 2.5\,{\rm{Hz}}$, and $\Gamma/2\pi  = 5.75\,{\rm{MHz}}$.}
	\label{diffSweepRate}
\end{figure}
\twocolumngrid
For larger sweep rates ($>\!\! 10$~$\mu$G/s), the asymmetry increases until the signal starts to oscillate [Fig.~\ref{diffSweepRate}(b) $\&$ (c)], while crossing $B=0$ \footnote{It should be noted that in the considered case [Fig.~\ref{diffSweepRate}(c)] the oscillations are also observed for negative fields. This effect, however, is an artifact of our experimental procedure where magnetic-field is scanned with a generator using a triangular waveform. Due to experimental limitations, in this case, the initially negative voltage reverses its direction and starts rising before the oscillations of first half-cycle of magnetic field sweep are completely damped. In that case, the oscillating component survives and is observed during the initial part of the dependence during rising, yet negative magnetic fields.}. The oscillations indicate that the system undergoes changes much faster then its relaxation rate. In such a case, experimental parameters (magnetic field) change significantly before equilibration so that atoms continuously tend toward a different equilibrium. As the parameters change continuously, this leads to time variation of the NMOR signal and appearance of the oscillations. The oscillations are damped and their period rises due to continuous increase of the magnetic-field strength ($B > 0$). The damping is a result of coherence relaxation through a reservoir effect (collisions of atoms with uncoated cell surfaces, particularly, metallic droplet within the stem) but also repumping of atoms with CW light (orientations of spins generated at different times vary, which leads to averaging transverse polarization, while longitudinal polarization is not affected or even enhanced). Since these processes depend on time not on a instantaneous value of magnetic field, the signal decay rate remains constant for different sweep rates. 
  
In Fig.~\ref{diffSweepRate}(a), at low sweep rate, amplitude of the signal is roughly a factor of two smaller than at the higher sweep rate [Figs.~\ref{diffSweepRate}(b) $\&$ (c)]. We checked in simulations that the increase in rotation angle at higher sweep rates is due to the presence of residual field in our experiments (more details will be presented in Ref.~\citep{RSGrewal}). In scope of this observation, we have introduce some residual magnetic field to our simulations [Figs.~\ref{diffSweepRate}(d) to (f)] to better reproduce the experimental conditions.  

The polarization rotation for a given sweep rate (914~$\rm{\mu}$G/s) but various light intensities is next investigated (Fig.~\ref{differentlight}). At low light intensities [Fig.~\ref{differentlight}(a)], the passage of magnetic field through zero causes oscillations of NMOR signal. The behavior is similar to that demonstrated in Fig.~\ref{diffSweepRate}(c). At higher intensities the oscillations are observed but they decay faster [Fig.~\ref{differentlight}(b)]. No oscillations are observed at even higher light intensities [Fig.~\ref{differentlight}(c)]. The reason for such a change in the observed dynamics is light-induced modification of the ground-state relaxation rate (power broadening), which causes that the system reaches the equilibrium state faster. In turn, at a given rate the system may equlibrate before physical conditions are significantly changed, or not equilibrate if the relaxation is long. Thereby, the oscillations of the NMOR signal observed at higher light power (larger power broadening) disappears and a signal similar to that traditionally recorded in NMOR is observed.

To demonstrate that the sweep rate not the magnetic-field scanning frequency or the change amplitude determines the transient dynamics of NMOR signal, the polarization rotation is measured at different scanned-field amplitudes and frequencies but same sweep rates (Fig.~\ref{samesweeprate}). The data shown in Fig.~\ref{differentlight}(a) confirm that despite the difference in scanning-field frequency and amplitude, the same transient behavior is observed in both signals for a given sweep rate.  Particularly, oscillation amplitude and time period is independent from individual values of both these parameters.

\onecolumngrid

\begin{figure}[H]
\centering  
\includegraphics[height=10.2cm,width=16.8cm]{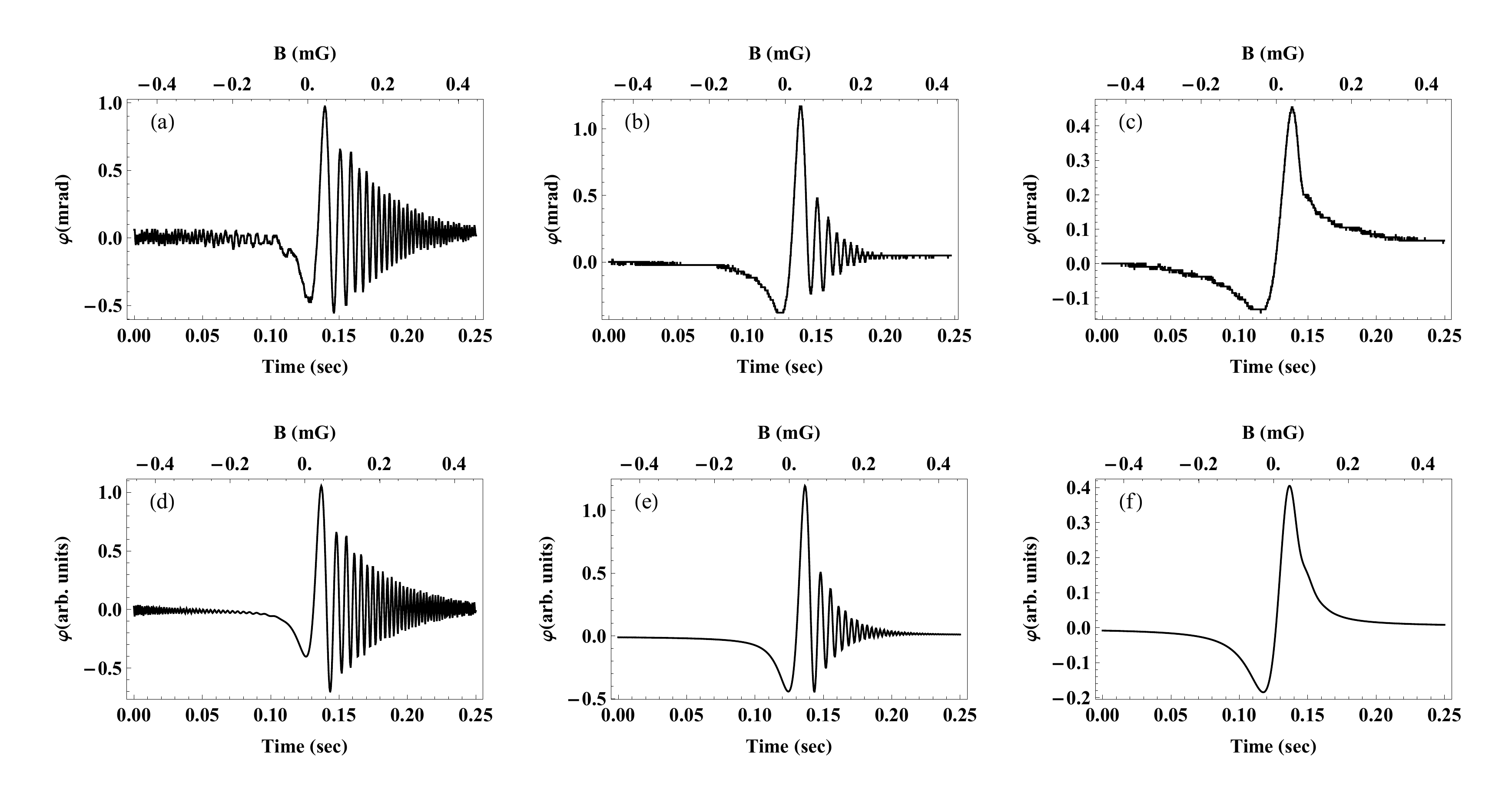}
\caption{Experimentally measured (top row) and numerically simulated (lower row) NMOR signals as a function of time  and magnetic field with magnetic field sweep rate  914 $\rm{\mu G/s}$ at different light intensities, (a) 0.71 $\rm{\mu W/mm^2}$ (b)  4.2 $\rm{\mu W/mm^2}$ (c)  21.2 $\rm{\mu W/mm^2}$, and Rabi frequencies (${\Omega _R}/2\pi)$, (d)  $7\,{\rm{kHz}}$, (e)  $  16\,{\rm{kHz}}$, and (f)  $  46\,{\rm{kHz}}$ (other parameters are same as before). The results show that damping of the oscillations depends on light intensities, and proves that the damping is a manifestation of faster relaxation due to power broadening and polarization repumping}
\label{differentlight}
\end{figure}
\twocolumngrid

In order to supplement the experimental data, the NMOR signals are simulated using the approach described in Sec.~\ref{sec:Modeling}. The signals are calculated for a set of parameters identical to that used in the experiments except the Rabi frequency. The difference in the Rabi frequency stems from simplifications of our model. Particularly, in our model atoms are motionless while in the real experiment they continuously go in and out of the light beam. As shown in Ref.~\citep{Zhivun2016Light}, this motion results in spatial averaging of light power over whole cell volume and hence effective lowering of the Rabi frequency. In turn, the lower value of Rabi frequency more thoroughly corresponds to the experimental situation than the value directly extracted from light intensity.

\begin{figure}[h]
\centering  
\includegraphics[width=\columnwidth]{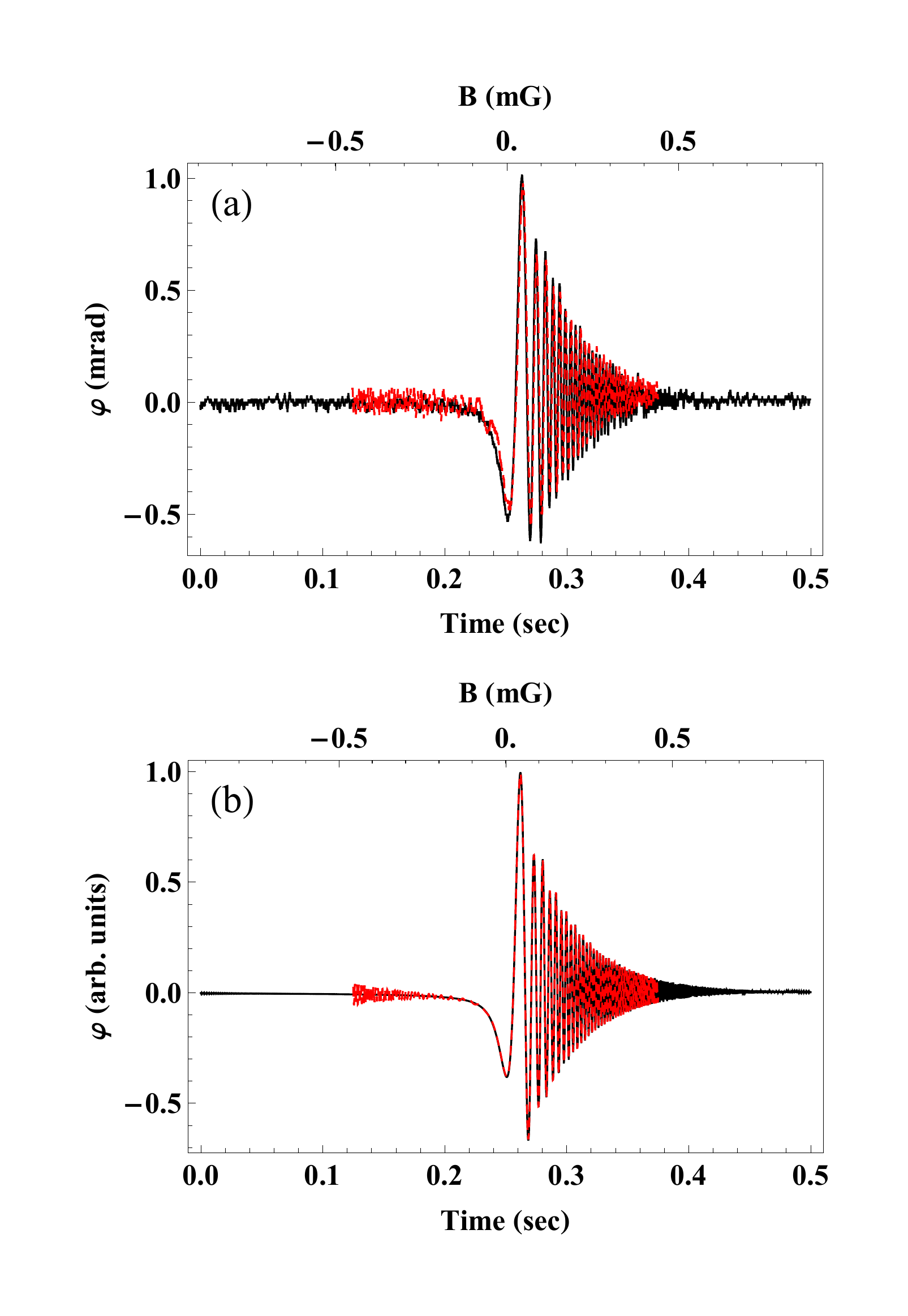}
\caption{Experimentally measured (a) and numerically simulated (b) NMOR signals as a function of time at a magnetic-field sweep rate of 914 $\rm{\mu G/s}$.  The dashed curve (red) and  solid curve (black) show the rotation signals for ${B_{0}}$ = 457 $\rm{\mu G}$ and $\Omega_{n}/2\pi=2~ $Hz, and for  ${B_{0}}$=914 $\rm{\mu G}$ and $\Omega_{n}/2\pi=$ 1 Hz, respectively. The measurements are performed with light intensity, (a) 0.71 $\rm{\mu W/mm^2}$, and (b) Rabi frequency ${\Omega _R}/2\pi =   7\,{\rm{kHz}}$.}
\label{samesweeprate}
\end{figure}

First, the time-dependent NMOR signals are computed at different magnetic-field sweep rate. As shown in Fig.~\ref{diffSweepRate}, a good agreement between experimental data and simulations results is achieved. Particularly, for low sweep rates, the (quasi-)symmetric NMOR signal is observed [Fig.~\ref{diffSweepRate}(d)]. This behavior changes when sweep rate is increased, i.e., when damped oscillations arise [Figs.~\ref{diffSweepRate}(e) $\&$ (f)]. We also reproduce the rotation signals dependence on light intensity by modifying the Rabi frequency (Fig.~\ref{differentlight}). Finally, we confirm that the signal dynamics depends on the sweep rate, not on scanning field amplitude or frequency [Fig.~\ref{samesweeprate}(b)]. 

To further investigate the transient dynamic of the system and to derive quantitative information, we adapt an approach developed in Ref.~\citep{Valente} for electromagnetically induced transparency and absorption. In the approach, the density matrix $\rho$ is reorganized into the vector $\vect{R}$ and the Liouville equation is rewritten into the form
\begin{equation}
\dot{\vect{R}}=M\vect{R}+\vect{R}_0,
\label{eq:EquationY}
\end{equation}
where $M$ is the matrix describing all density-matrix-dependent processes and $\vect{R}_0$ is the vector describing all the density-matrix-independent processes. Via diagonalization of the matrix $M$, one can write the solution of Eq.~\eqref{eq:EquationY} in a general form
\begin{equation}
	\vect{R}=\sum_i a_i\vect{v}_i\exp{(\lambda_it)}-M^{-1}\vect{R}_0,
	\label{eq:SolutionY}
\end{equation}
where $\lambda_i$ are eigenvalues and $\vect{v}_i$ eigenvectors of the matrix $M$, and the coefficients $a_i$ depend on the initial state of the system. 

\begin{figure}[!h]
	\includegraphics[width=\columnwidth]{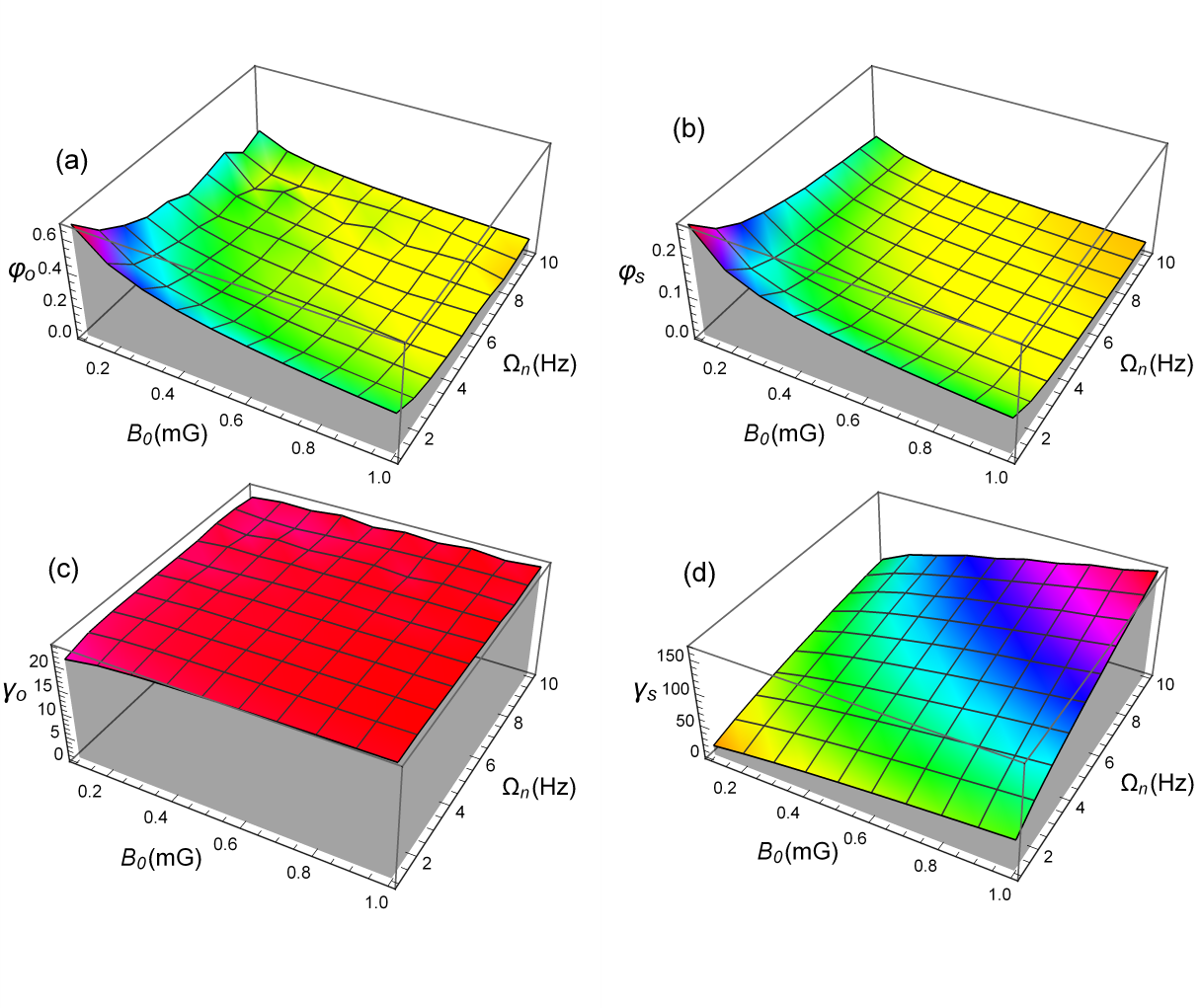}
	\caption{Amplitude $\varphi_o$ (a) and decay rate $\gamma_o$ (b) of the oscillations component and amplitude $\varphi_s$ (c) and  decay rate $\gamma_s$ (d) of the static component extracted using Eq.~\eqref{eq:Fitting} as functions of scanning field frequencies and amplitudes. The parameters used in simulations are, ${\Omega _R}/2\pi = 7\,{\rm{lHz}}$, $\gamma/2\pi  = 2.5\,{\rm{Hz}}$, and $\Gamma/2\pi  = 5.75\,{\rm{MHz}}$. }
	\label{3dplot}
\end{figure}

In the case of two-level atom, one can use Eq.~\eqref{eq:SolutionY} to write the equation for the time dynamics of our system
\begin{equation}
	\varphi=\varphi_o\exp[-\gamma_o (t-t_0)]\sin[S (t-t_0)^2+\phi]+ \varphi_s \exp[-\gamma_s (t-t_0)],
	\label{eq:Fitting}
\end{equation}
where $\varphi_o$ and $\varphi_s$ are the amplitudes of the oscillatory and static part of the signals, respectively, $\gamma_o$ and $\gamma_s$ are corresponding decay rates, $\phi$ is the oscillation initial phase, $S$ is the sweep rate, given in Hz/s units, and $t_0$ is the zero-crossing time. 

Equation~\eqref{eq:Fitting} reveals existence of two components in the observed signal. The first is the oscillatory component, arising due to nonequilibrium dynamic of the system. It is related to the ground-state coherences and its evolution in the magnetic field. The second component does not reveal any oscillations. It is believed that the component corresponds to the equilibrium processes, i.e., the processes arising at faster time scale. An example of such a process is optical pumping.

Besides the amplitude and relaxation rates, Eq.~\eqref{eq:Fitting} also contains the phase term $\phi$, which accounts for existence of nonnegligible Zeeman coherence at $t_0$. While such a coherence may be generated at $t<t_0$, the efficiency of the process depends on the sweeping rate (the faster rate, the less time for coherence generation). Hence there is nonzero coherence that contributes to the coherence at $t>t_0$. Such a coherence may also arise as an artifact of our measurement procedure and implementation of continuous scanning of magnetic field with a triangular waveform. For higher scanning frequencies (shorter scanning periods), the coherences may not completely relax before entering successive sweep cycle. Both effects may cause oscillations at $t<t_0$, thus the phase term, and sweep-rate-dependent modification of the oscillation amplitude (see below).

To study the dependence of the NMOR signal on the magnetic-field sweep rate, we simulate the rotation signals at magnetic-field sweep amplitudes (from 0.1 mG to 1.0 mG) and frequencies (from 1~Hz and 10~Hz) and next fit the signals according to Eq.~\eqref{eq:Fitting}. Figure~\ref{3dplot}(a) presents the dependence of the oscillation amplitude $\varphi_o$ on both parameters. As shown, the amplitude decays with the sweep-rate increase (either through increase of the magnetic-field scanning amplitude or frequency). This is caused by the fact that faster sweepings leads to less efficient coherence generation (there is less time for medium polarization). Interestingly, at higher scanning frequencies,  some oscillations in $\varphi_o$ are observed. For example, there is a significant increase in the rate for $\Omega_n = 8$~Hz. 

To investigate this process, we simulate the oscillation amplitude $\varphi_o$ as a function of scanned magnetic-field amplitude, fixed modulation frequency, and two distinct Rabi frequencies (Fig.~\ref{fig:amplitude}). The amplitude of  oscillatory component decreases with the modulated-field amplitude but additional modulation in the dependence is observed. A period of the modulation is independent from the Rabi frequency, but its amplitude strongly depends on the quantity.
\begin{figure}[htb!]
	\includegraphics[width=\columnwidth]{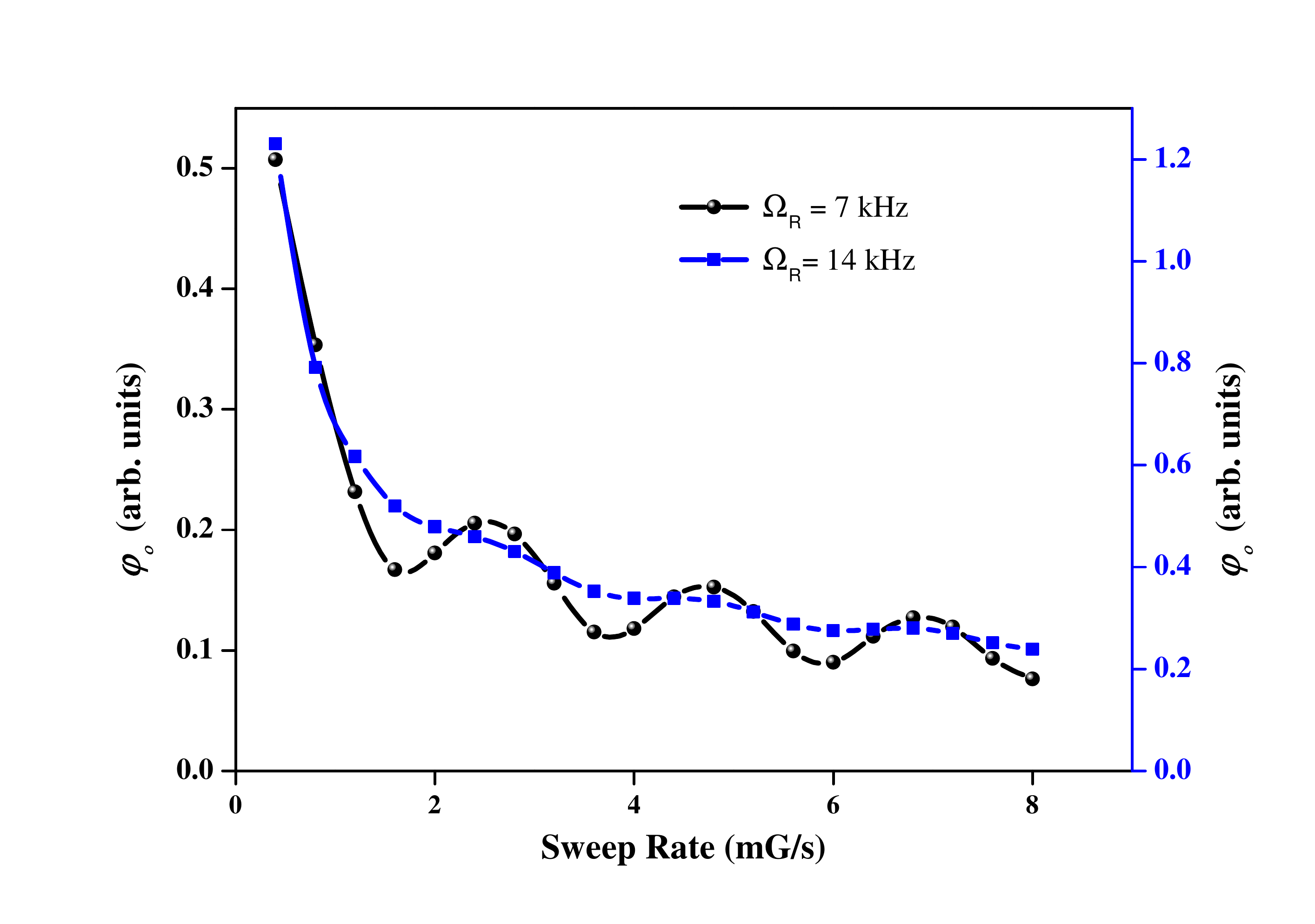}
	\caption{Amplitude of the oscillatory component $\varphi_o$ of the observed NMOR signal versus the sweep rate for two Rabi frequencies at scanning frequency $\Omega_n$ = 8 Hz. The modulation of the amplitude, observed for much weaker Rabi frequency, is the effect of the interference between initially created coherence  $t<t_0$ and the coherences generated for $t>t_0$. If both coherences are in phase, amplitude of the signal is enhanced. Whereas, if they are out of phase, it is deteriorated.}
	\label{fig:amplitude}
\end{figure} 
The oscillations observed in the $\varphi_o$ dependence stem from optical pumping at $t<t_0$. This earlier generated Zeeman coherence evolve due to magnetic field and they sum up with the coherences generated at $t>t_0$. Depending on the dynamics of the coherence evolution, determined by the sweep rate, the coherences may add up constructively (if the precession phase matches) or destructively (when the coherences add out of phase). The addition of the coherences results in enhancement or deterioration of the rotation. Besides simple phasing of the coherences generated at different times, one should also considered relaxation of the coherences. In general, the faster the relaxation, the less ``memory'' the medium has and hence less pronounced the intereference effect is. This is why, for larger light intensities the effect is less significant (larger ground-state relaxation) than for less intense light (when relaxation is prolonged).

Figure~\ref{3dplot} also shows three other fitting parameters: amplitude of the nonoscillation component $\varphi_s$, and the decay rates of the oscillation and nonoscillating components, $\gamma_o$ and $\gamma_s$, respectively. While the amplitude dependence of the nonoscillating component follows the same trend as oscillatory-component amplitude, no modulation in the dependence is observed and the nonoscillating component value is three times smaller than the amplitude of the oscillating component. The decay rate of the oscillating component $\gamma_o$ is shown in Fig.~\ref{3dplot}(c). Interestingly, the rate does not depend on magnetic-field scanning parameters (small fluctuations arise due to nonideal fitting). This proves that relaxation of the oscillating component is predominantly determined by the reservoir effect. This contrasts the dependence of the decay rate of the static component $\gamma_s$, shown in Fig.~\ref{3dplot}(d), which strongly depends on the sweep rate. The different behavior indicates that the observed static rotation explicitly depends on the contemporary magnetic field, but the system equilibrate much faster so that no rotation is observed. This observation suggest that the stationary part may be related to optical pumping.
 
Numerical simulations provides additional capabilities in investigating dynamics of the system. For example, one can investigate NMOR signals on the ground-state relaxation rate and Rabi frequency, independently. Such deconvolution is not possible in our experiment where changing the Rabi frequency simultaneously changes the ground-state relaxation rate.

 \begin{figure}[H]
	\includegraphics[width=\columnwidth]{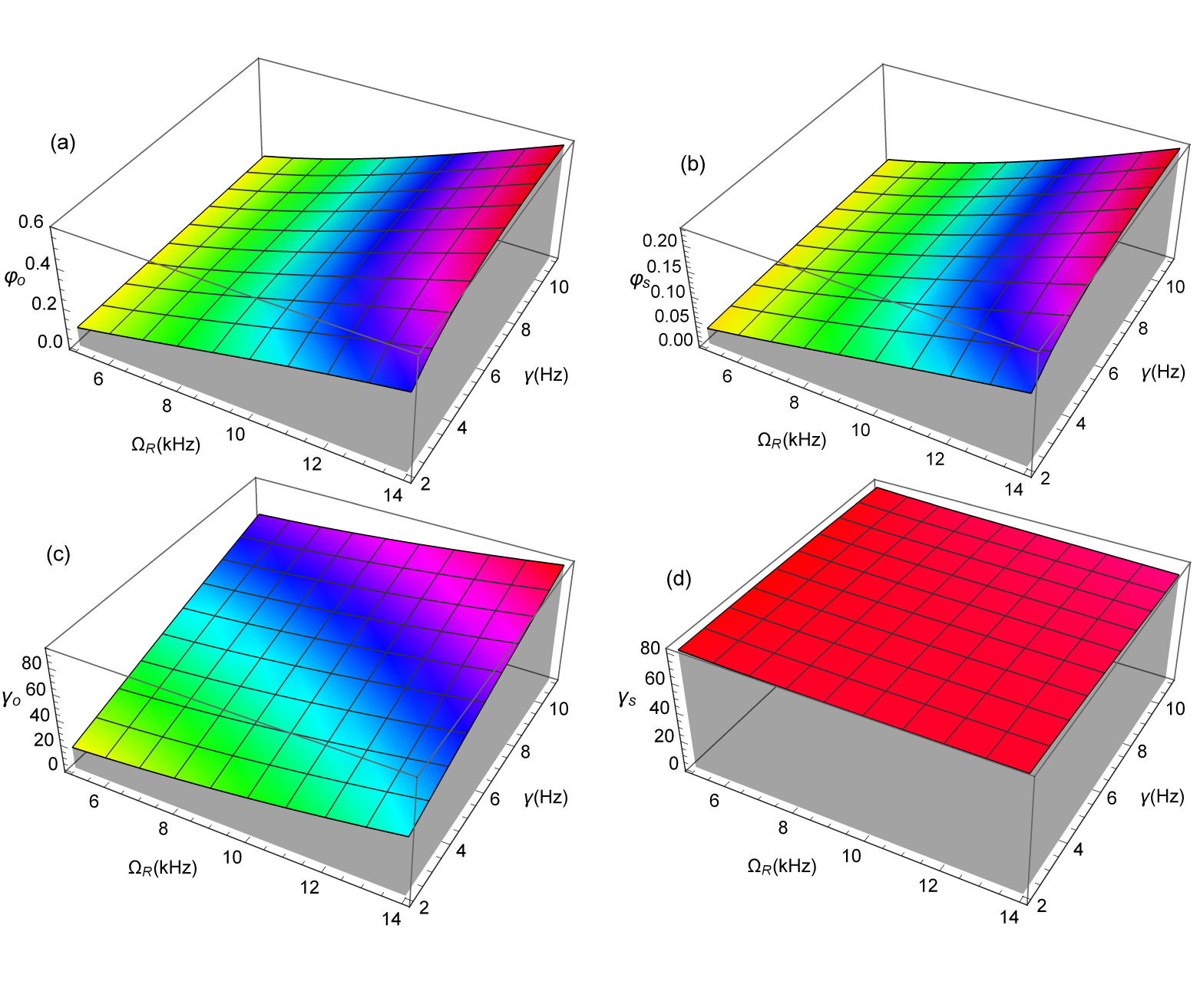}
	\caption{Amplitude $\varphi_o$ (a) and decay rate $\gamma_0$ (b) of the oscillating component of the signal and amplitude $\varphi_s$ (c) and decay rate $\gamma_s$ (d) of the static component of the observed signals as functions of the Rabi frequency $\Omega_R$ and ground-state relaxation rate $\gamma$ at a given sweep rate (2.5~mG/s).}
	\label{3dplot2}
\end{figure}

To fully explore the dynamics of the system, the NMOR signals are simulated for different Rabi frequencies and ground-state relaxation rates at a fixed sweep rate. Figure~\ref{3dplot2} shows the dependence of the amplitude $\varphi_o$ and decay rate $\gamma_o$ of the oscillating component, as well as the amplitude $\varphi_s$ and decay rate $\gamma_s$ of the static component on both parameters. Similarly as before, the amplitudes of oscillating and static components reveals analogous dependences on both parameters [Figs.~\ref{3dplot2}(a) $\&$ (b)]. For instance, amplitude of both components increase when more-intense light is used. This suggests that coherence are generated more efficiently at higher Rabi frequencies. Somewhat counter-intuitively, the dependences also show that at higher Rabi frequencies, the amplitudes increase with ground-state relaxation. This effect may be explained as follows. For higher Rabi frequencies the system is saturated. The saturation means strong repumping of initially created coherence, which leads to deterioration of the signal. For higher ground-state relaxation the saturation is reduced and the effect is limited, leading to signal increase. Besides amplitudes, Fig.~\ref{3dplot2} also presents the decay rate of the oscillating component $\gamma_o$. The rate increases with both the ground-state relaxation rate $\gamma$ and the Rabi frequency $\Omega_R$. While the former affects the rate directly, the latter effects the rate via power broadening. It is noteworthy that, however the scaling of $\gamma_o$ with the ground-state rate $\gamma$ is linear, but it is not one-to-one. In contrast to $\gamma_o$, the static decay rate $\gamma_s$ is almost independent on $\gamma$ and $\Omega_R$. Such a dependence suggests that the static and oscillating components are having different source of origin. We believe that optical pumping is responsible for the process.

\section{Conclusion}

We have studied the transient dynamics of nonlinear magneto-optical rotation in a paraffin-coated Rb cell at weak yet continuously varying magnetic field. We demonstrated that at low magnetic-field sweep rate, the shape of the rotation signal is similar to the traditionally observed NMOR signal. The only difference is a small asymmetry between two wings of the signals, which is the first indication of transient response of the atomic medium. When the sweep rate is increased, the signal starts to oscillate during crossing zero field. The origin of these oscillations is well understood and it indicates that the atomic medium does not achieve equilibrium before significant change of physical parameters occurs. At the same time, the oscillations die out due to relaxation, averaging with optical pumping during the spin precession. We further investigated the effect of scanning magnetic-field amplitude and frequency on the transient response. We demonstrated that the transient response of two signals with equal sweep rate but different scanning amplitudes and frequencies is the same. The dependence on the ground-state relaxation rate is investigated by increasing the light intensity (experiment). It is shown that for the higher intensities, the system reaches equilibrium faster and no oscillations are observed. This is also confirmed with numerical calculations where the Rabi frequency and ground-state relaxation rate are independently modified.

The proposed scheme may enable detection of transient magnetic field or other scalar spin couplings. The effect may be used to detected abrupt changes of magnetic field such as induced by rapid demagnetization of materials or pulses of currents (electric sparks). In fundamental research, the effect may be used in searches for transient exotic spin couplings caused by axion-like-particle topological defects \citep{Pospelov2013Detecting} performed with the Global Network of Optical Magnetometers for Exotic physics searches \citep{Pustelny2013Global}.

\begin{acknowledgments}

The authors would like to acknowledge support from the National Centre of Research and Development within the Leader programme (RSG) and the National Science Centre with the Opus programme (SP).

\end{acknowledgments}


\nocite{*}
\bibliography{transientNMOR}

\end{document}